\documentclass[preprint,amssymb]{revtex4-1} 
\usepackage{amsmath,amssymb,latexsym,graphicx,color,hyperref,subfig, graphicx}
\newcommand{\bx}{\mbox{\bf x}}
\newcommand{\by}{\mbox{\bf y}}

\newcommand{\bg}{\mbox{\bf g}}

\newcommand{\bu}{\mbox{\bf u}}

\newcommand{\FF}{\mbox{\bf F}}

\newcommand{\ff}{\mbox{\bf f}}

\begin{document}

\title{Elastohydrodynamics of swimming helices: effects of flexibility and 
       confinement}

\author{John Lagrone}
\email{jlagrone@tulane.edu}
\affiliation{Department of Mathematics, Tulane University, New Orleans, 
             LA 70118}

\author{Ricardo Cortez}
\email{rcortez@tulane.edu}
\affiliation{Department of Mathematics, Tulane University, New Orleans, 
             LA 70118}


\author{Lisa Fauci}
\email{fauci@tulane.edu}
\affiliation{Department of Mathematics, Tulane University, New Orleans, 
             LA 70118}

\date{\today}

\begin{abstract}
Motivated by bacterial transport through porous media, here we study the
  swimming of an actuated, flexible helical filament in both three-dimensional
  free space and within a cylindrical tube whose diameter is much
  smaller than the length of the helix.  
The filament, at rest, has a native helical shape modeled after the geometry of
  a typical bacterial flagellar bundle.  
The finite length filament is a free swimmer, and is driven by an applied 
  torque as well as a counter-torque (of equal strength and opposite direction)
  that represents a virtual cell body.  
We use a regularized Stokeslet framework to examine the shape changes of the
  flexible filament in response to the actuation as well as the swimming
  performance as a function of the nondimensional {\em Sperm number} that 
  characterizes the elastohydrodynamic system.  
We also show that a modified Sperm number may be defined to characterize the
  swimming progression within a tube. 
Finally, we demonstrate that a helical filament whose axis is not aligned with
  the tube axis can exhibit centering behavior in the narrowest tubes. 
\end{abstract}

\maketitle

\section{Introduction}

At the microscale, bacterial motility is achieved through the action of 
  rotating helices \cite {lauga2016bacterial}.  
Monotrichous cells such as {\em Pseudomonas aeruginosa} are propelled by a
  single helical flagellum driven by a rotary motor at its base.  
Peritrichous cells such as {\em Escherichia coli} are propelled by many helical
  flagella that, when rotating in the same direction, form a coherent helical
  bundle.  
A fascinating family of bacteria,  spirochetes, have cell bodies that are
  themselves helical and appear to move as a corkscrew through viscous fluids 
  \cite {charongoldstein}.
In addition to nature's swimming helices, fabricated helical micromachines
  present intriguing possibilities in biomedical applications such as drug 
  delivery \cite{nano2009,pham2015}.  
While varying in their material properties, these helices (individual bacterial
  flagella, flagellar bundles, spirochete cells, fabricated helices) are elastic 
  structures that, when actuated in a viscous fluid, could experience
  shape deformation in response to the flow.    

It is also of interest, both for engineered helical microswimmers and natural
  bacterial cells, to understand how motility is affected when moving through 
  confined environments.  
Indeed, the use of bacteria in bioremediation often relies on cells moving 
  through porous media such as soil, where the pore sizes are smaller than the
  flagellar length \cite{ping2015}.  
The effects of moving through such restrictive geometries on the swimming
  performance of bacterial cells has been examined both in laboratory experiments 
  and mathematical models. 
Early experiments \cite{kyr1995} of {\em E. coli} swimming in glass capillary
  tubes of three micron diameters showed that the cells could not tumble and
  reorient in this confined space, and exhibited unidirectional motion. 
More recently, it was shown that such tumbling of {\em E. coli} was also 
  hindered when cells were swimming close to a planar surface.  
Microfluidic experiments showed that {\em E. coli} cells tend to swim in 
  helical paths in narrow capillaries \cite{ping2015}, and {\em S. marcescens}
  experienced a sizable boost in swimming velocity in microchannels 
  \cite{binz2010}.

Models of microbial swimming in cylindrical capillary tubes study the
  performance of spherical squirmers \cite{zhu2013low} and dipolar swimmers
  \cite{yeomans2013}.
Using a boundary element method, Zhu et al. \cite{zhu2013low} show that
  spherical squirmers with tangential deformations swim more slowly as
  confinement increases, but swimmers with normal deformations swim faster with
  confinement.  
The reduced model swimmers in \cite{yeomans2013} moving in the center of a
  rigid tube move faster with confinement.   
A model of a swimmer in a capillary tube that represents the detailed geometry
  of a rotating rigid helix was presented by Liu et al. \cite{Liu2014}. 
In this model, both the tube and the helix were infinitely long, but the radius
  of the tube was on the order of the helical radius.  
Except for the tightest confinements, swimming speed increased with confinement
  when a fixed torque was applied.  

As in the model of \cite {Liu2014}, here we study the swimming of an actuated
  helical filament in a capillary tube whose radius is close to the helix
  radius.
However, the helical filament in this study is flexible and of finite length.  
Using a regularized Stokeslet framework  \cite{cortez2005method}, we first
   examine the swimmer in free space, and show that its performance is
   well-characterized by the Sperm number, a non-dimensional parameter that
   measures the ratio of viscous fluid forces to elastic forces.   
We then examine the effects of confinement in a tube, and suggest a 
  modification of  the Sperm number to account for the effect of the tube
  surface.
Finally, because the helical filament is of finite length, we can study its
  dynamics when it is initialized at an angle to the centerline of the tube.

\section{Methodology} \label{methods}

\subsection{Stokes equations}

We model a flexible, helical filament that is  actuated by applied torques in a
  viscous fluid, where the length and time scales are small enough that
  inertial forces are negligible. 
The fluid motion is therefore well-modeled by the incompressible Stokes
  equations:
\begin{align}
 \begin{split}
	0 &= -\nabla \hat{P} + \mu \Delta \hat{\mathbf{u}} + \hat{\mathbf{F}} +
	  \frac{1}{2} \nabla \times \hat{\mathbf{L}}, \\
	0 &= \nabla \cdot \hat{\mathbf{u}}, 
    \label{stokes}
 \end{split}
\end{align}
where $\hat{P}$ is the pressure, $\hat{\mathbf{u}}$ is the fluid velocity,
  $\mu$ is the fluid viscosity, $\hat{\mathbf{F}}$ is the external force per
  volume and $\hat{\mathbf{L}}$ is external torque per volume exerted by the
  helical swimmer on the fluid.

The forces and torques in Eqn. \ref{stokes} will be localized at the helical
  filament.   
These equations hold in all of three-dimensional space.  
To nondimensionalize the problem we assume characteristic scales for length
  $\hat\ell$,  time $\hat{T}$, force ${\hat{\mathcal F}}$, and  torque
  ${\hat{\mathcal L}}$.
By choosing ${\mathcal F}=\mu \hat\ell^2/\hat{T}$ and 
  ${\mathcal L}=\mu \hat\ell^3/\hat{T}$, the  dimensionless Stokes equations
  are:
\[
0 = - \nabla p +\Delta\bu + \FF + \frac{1}{2} \nabla \times \mathbf{L}
\]
\noindent
Throughout this manuscript, we choose the characteristic scales to be: 
  $\hat\ell = $ 4 $\mu$m,  
  $\hat{T} = $ 0.01 sec,
  $\mu$ to be the viscosity of water, and, hence, $\cal F$ = 
  $1.6\times10^{-12}$ N.

We  use a regularized Stokeslet framework \cite{cortez2005method} to model the
  elastohydrodynamic system, where the external force comes from a surface
  integral of regularized forces supported on the cylindrical surface of the
  helical filament,  while the regularized torques are applied only at two
  points $\by_1$ and $\by_2$.  
The first torque will be applied at the tip of the helix, whereas the second,
  of equal strength  but opposite direction, will be applied slightly in front
  of the filament, as a proxy to a counter-rotating cell body.  
The expressions for force and torque are:   
\[
  \FF(\bx) = \int_\Sigma \ff(\by) \phi_{\epsilon}(\bx-\by) dS_y,\ \ \ 
  \mathbf{L}(\bx) = \sum_{k=1}^2 \bg_k \phi_{\epsilon}(\bx-\by_k).
\]
The regularization (or blob) function is chosen to be:
\begin{align}
 \phi_\varepsilon (\mathbf{x} - \mathbf{y}) 
 = \frac{15 \varepsilon^4}{8 \pi (r^2 + \varepsilon^2)^{7/2}},
\end{align}
where $r = \| \mathbf{x} - \mathbf{y} \|$.
This leads to the velocities due to the regularized Stokeslets and rotlets as
  follows: 
\begin{align} \label{eq:stokelet}
  {\mathbf{u}}_{st}(\mathbf{x}) = \int_\Sigma
  S_\varepsilon(\mathbf{x},\mathbf{y}) {\mathbf{f}}(\by) dS_y
  = \frac{1}{8 \pi} \int_\Sigma \frac{(r^2 + 2 \varepsilon^2)
  {\mathbf{f}}(\by)  + ({\mathbf{f}}(\by) \cdot (\mathbf{x} - \mathbf{y}))
  (\mathbf{x} - \mathbf{y})}{(r^2 + \varepsilon^2)^{3/2}} dS_y. 
\end{align}
\begin{align} \label{eq:rotlet}
 {\mathbf{u}}_{rt}(\mathbf{x}) = \sum_{k=1}^{2}
 R_\varepsilon(\mathbf{x},\mathbf{y}_k) \mathbf{g}_k = \frac{1}{16 \pi}
 \sum_{k=1}^{2} \frac{2r_k^2 + 5 \varepsilon^2}{(r_k^2 + \varepsilon^2)^{5/2}}
 \left(\mathbf{g}_k \times (\mathbf{x} - \mathbf{y}_k) \right),
\end{align}
where ${\mathbf{f}}(\by)$ is force per unit area, $\Sigma$ denotes the surface
  of the helical filament, $\bg_k$ is torque, $r=|\bx-\by|$, $r_k=|\bx-\by_k|$
  and $\varepsilon$ is the regularization parameter.
We note that these velocities are defined everywhere in $\mathbb{R}^3$ and are
  everywhere incompressible.

\subsection{Representation of helical filament and its actuation}

The model elastic filament that we consider has a native helical equilibrium
 shape whose centerline is given by:
\begin{align}
 \begin{split}
    x({s}) & = \alpha({s}), \\
    y({s}) & = -{r}_h({s}) \cos \left(\frac{(2 \pi n_p)  {s}}{{L}}\right),
               \qquad 0  \leq s \leq {L}, \\
    z({s}) & = {r}_h({s}) \sin \left( \frac{(2 \pi n_p) {s}}{{L}}\right), 
 \end{split}
\end{align}
where ${L}$ is the arc length of the helix, $n_p$ is the number of turns in the
  helix, and the helical radius, tapered from back to front, is given by:
\begin{align}
  {r}_h({s}) = {A} \left[ \frac{1}{\pi} \arctan 
  \left(\frac{{\beta s}}{{L}} - 1 \right) + \frac{1}{2}\right].
\end{align}
As in \cite{Flores2005}, $\alpha({s})$ is chosen such that the tangent vector
  $[x'({s}), y'({s}), z'({s})]$ has unit length, so that $s$ is an arclength
  parameter.  
We do not view the helical filament as a single bacterial flagellum, but rather
  as a representation of a bacterial flagellar bundle.  
In this work, we choose a fixed equilibrium configuration of the helical 
  filament in all simulations, whose associated geometric parameters are given
  in Table \ref{bigtable}.
These fall within the range of parameters for a typical, loosely packed helical
  bundle \cite {lauga2016bacterial, Turner2000, macnab1977bacterial}.

\begin{table}
  \begin{center}
    \begin{tabular}{|l|c|c|}
      \hline
      Input Quantities & Dimensionless & Corresponding  \\
      &  value  & dimensional value \\
      \hline
      Helix arc length, $L$  & 5.42   & 21.7 $\mu$m \\
      Helix projected length, $l$   & 3.93    &  15.7     $\mu$m \\
      Number of pitches, $n_p$    & 2 & 2 \\
      Helix max amplitude, $A$ &  0.5       &  2 $\mu$m       \\
      Tapering parameter, $\beta$        & 6 & 6  \\
      Filament radius, $R_f$             &0.08& 0.32 $\mu$m \\
      Rotlet strength $\sigma$  (= 0.5 torque) & $1.0 - 7.0$  & $(6.4 - 44.8)
         \times 10^{-18}$ N-m   \\
      Spring stiffness, $k$   & $75 - 1200$  & $ 1.2 \times 10 ^{-10} - 1.92
         \times 10^{-9}$  N \\
      $EI$   & $2.36 - 37.1$  &  $(6.0 - 96.5)\times 10^{-23}$ N-m$^2$    \\
      Counter rotlet separation, $\tau$    &  $.50 $ & $2 \mu$m\\
      \hline
      Computed Quantities &  &   \\
      \hline
      Frequency, $\omega$        & 0.024 - 0.67    & 2.4 - 67 Hz    \\
      Swimming speed, $U$  &  0.0057 - 0.14     &  2.28 - 55.5 $\mu$m/s     \\
      Distance per revolution & 0.05 - 0.34 & 0.20 - 1.4 $\mu$m    \\
      \hline
      Numerical Parameters &  &  \\
      \hline
      Cross sections along helix, $N_f$     & 65     & 65  \\
      Points per helical cross section, $N_c$       & 6        & 6  \\
      Cross sections along tube, $N_t$    & 64         &64  \\
      Points per tube cross section, $N_{\theta}$   & 24 - 38      & 24 - 38  \\
      Spacing between helix nodes, $\Delta s_h$ & $ 0.08$  & 0.32 $\mu$m \\
      Helix blob size, $\epsilon_{h}$ &0.025  & 0.1 $\mu$m \\
      Spacing between tube nodes, $\Delta s_{t}$ &  0.16       &0.64 $\mu$m  \\
      Tube blob size $\epsilon_{t}$ & 0.5 &0.2 $\mu$m \\
      Time step, $\Delta t$   & $2.5 \times 10^{-6}  - 1.0 \times 10^{-5}$ & 
        2.5 $\times 10^{-8}$s  - 1.0 $\times 10^{-7}$s\\ 
      \hline
    \end{tabular}
  \caption{Input parameters, computed quantities and numerical parameters used
  	       in computations.}
  \label{bigtable}
  \end{center}
\end{table}

\begin{figure}[h!tp]
  \centering
  \includegraphics[width=0.7\linewidth]{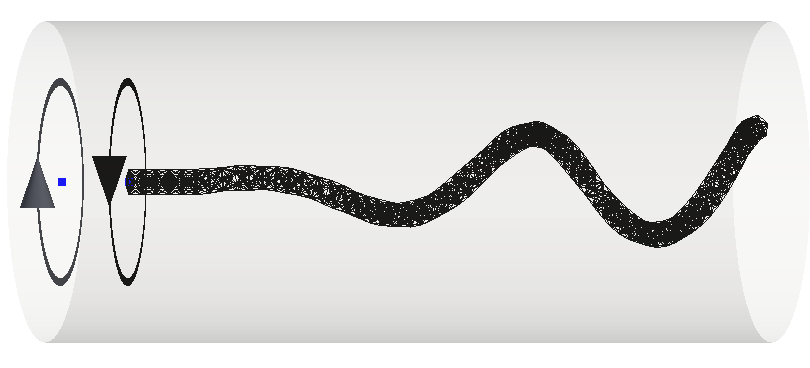}
  \caption{Computational helical filament consisting of a network of springs.
  	       The figure shows the equilibrium configuration. The motion is
  	       generated by the activation of a torque at the center of the front
  	       cross-section, and by the counter-torque which is placed a small
  	       distance in front of the first cross-section where the cell body
  	       might be.}
  \label{fig:flagella}
\end{figure}

We construct the discretization of the surface of the cylindrical helical
  filament by placing hexagonal cross-sections of radius $R_f$ along the
  helical centerline, perpendicular to the centerline. 
As such, each cross-section is discretized by $N_c$ = 6 six points, and we take
  $N_f$ cross-sections along the helical filament so that the spacing between
  neighboring cross-sections is approximately equal to the spacing between
  adjacent points on a cross section (see Figure \ref {fig:flagella}).  
Each of the $N = N_c \times N_f$ discrete  points on the surface of the 
  filament is connected to a subset of the other surface points by a Hookean
  spring, giving elasticity to the structure.  
We define the dimensionless elastic energy in the system as
\begin{align}\label{eq:Energy}
  {\cal E} = \frac{1}{2} \sum_j k_j  l_j 
             \left(\frac{\| \bx_{j_1}-\bx_{j_2} \|}{l_j}-1 \right)^2, 
\end{align}
where $k_j$ is the stiffness of a spring with resting length $l_j$ that 
  connects points $j_1$ and $j_2$.  
The sum is over all springs.
The force at $\bx_{j_1}$ is $\ff_{j_1}A_{j_1}$ where $A_{j_1}$ is the area of a
  patch of surface centered at $\bx_{j_1}$ in the discretization. 
We have that
\[
  \ff_{j_1}A_{j_1} = - \frac{\partial {\cal E}}{\partial \bx_{j_1} }.   
\]
Similar constructs of semi-flexible filaments using nodes with elastic linkages
 have been used to model diatom chains \cite{Fauci4} and bacterial flagella
 \cite {peskinlim2004,Flores2005}.  
In all simulations shown here, we choose a spring topology so that each point
  on a given cross-section is connected to every other point on that 
  cross-section, as well as to every other point on the two cross-sections
  adjacent to it.  
This means that each node is connected to $5 + 2 \times 6 = 17$ other nodes.  
In addition, in all simulations shown, the stiffness constant $k_j = k$ in 
  Eqn. \ref{eq:Energy}  is taken to be the same for all springs.  
The resting lengths of the springs, $l_j$ in Eqn. \ref {eq:Energy} do vary with
  $j$, and are computed during the construction of the helical surface.  
The initialized helical filament configuration is in its equilibrium state 
  (the total energy in Eqn. \ref{eq:Energy} is zero).     

As in bacterial flagella that are driven by a rotary motor at their base, we 
  actuate the helical filament by a regularized torque that is placed at the
  centroid of the first cross-section $\mathbf{y}_1$.  
We apply a counter-torque of the same strength and opposite direction in front
 of the first cross section, at a fixed distance away.  
In practice this is done by computing the outward normal to the first
  cross-section $\mathbf{n}$ and choosing 
  $\mathbf{y}_2 = \mathbf{y}_1 + \tau \mathbf{n}$, where $\tau$ is a positive
  parameter. 
Due to these applied torques, the flexible filament will depart from its 
  equilibrium shape as the network springs become stretched or compressed, 
  causing forces at the nodes to develop.  
The fluid velocity due to these elastic forces and the applied torques 
  is evaluated at each material point $\mathbf{x}_i, i = 1,...,N$ of the 
  helical surface, using  (Eq.~\eqref{eq:rotlet}) and a discrete version of 
  (Eq.~\eqref{eq:stokelet}): 
\[
  \bu(\mathbf{x}_i) = \sum_{k=1}^{2} 
  R_\varepsilon(\mathbf{x}_i,\mathbf{y}_k) \mathbf{g}_k
  + \sum_{j=1}^{N} S_\varepsilon(\mathbf{x}_i,\mathbf{x}_j) \ff_{j}A_{j}.
\]
For a swimmer in free space, this velocity is used to update the positions 
  of the nodes of the flexible filament, satisfying the no-slip boundary
  condition of Stokes flow. 
Note that since all forces arise from springs, and the torque driving the
  filament is balanced by a counter-torque representing the cell body, the
  sum of forces and torques are zero, and, hence, momentum and angular 
  momentum are conserved.  
The actuated flexible helix is a free swimmer.

\subsection{Coupled helix-tube system}

We wish to examine the swimming of the actuated helical filament described 
  above in a rigid, cylindrical tube whose radius $R$ is smaller than the 
  filament length, but whose length is long enough so that no end effects 
  on the fluid dynamics are present. 
The surface of the tube is discretized by $N_t$ circular cross-sections 
  with $N_{\theta}$ points each, for a total of 
  $N_{tube} = N_t \times N_{\theta}$ points. 
Regularized forces on these $N_{tube}$ points $\mathbf{z}_i$  will be 
  computed so that the no-slip (zero velocity) condition is satisfied.  
These forces, plus the elastic forces supported on the helical filament, 
  as well as the applied torques will determine the fluid velocity at
  any point in the tube, along with the filament's swimming progression.  
Here we describe the overall algorithm for evolving this coupled 
  helix-tube system: 

Given helix surface points and forces  
  $\mathbf{x}_j,\mathbf{f}_j, j = 1,...,N$ and torques 
  $\mathbf{g}_1, \mathbf{g}_2$: 
\begin{enumerate}
\item
Compute the velocities on the $N_{tube}$ tube surface points that are 
  induced by these forces and torques: 
\[
  \mathbf{\tilde{u}}(\mathbf{z}_i) = \sum_{k=1}^{2} 
  R_\varepsilon(\mathbf{z}_i,\mathbf{y}_k) \mathbf{g}_k
  + \sum_{j=1}^{N} S_\varepsilon(\mathbf{z}_i,\mathbf{x}_j) \ff_{j}A_{j}
\]
\item
Compute forces $\mathbf{h}_i$ that must be exerted on the tube points so
  that the velocity $\mathbf{\tilde{u}}$ is cancelled out there:
\[
  - \mathbf{\tilde{u}}(\mathbf{z}_{i})
  =
  \sum_{j=1}^{N_{tube}} S_\varepsilon 
  (\mathbf{z}_{i},\mathbf{z}_{j}) \mathbf{h}_{j} \hat{A}_{j}
\]
This is a $3N_{tube} \times 3N_{tube}$ linear system for the unknown 
  forces  $\mathbf{h}_{j}$. 
We note that the (dense) coefficient matrix depends upon the relative
  distances between the discrete nodes on the tube's surface which do
  not change in time.  
This allows us to a precompute its factorization once, even though 
  the system is solved at every time step.
\item
Finally, exploiting linearity of the Stokes equations, the velocities
  of the material points on the helical filament are:  
\[
  \bu(\mathbf{x}_i) = \sum_{k=1}^{2} 
  R_\varepsilon(\mathbf{x}_i,\mathbf{y}_k) \mathbf{g}_k
  + \sum_{j=1}^{N} S_\varepsilon(\mathbf{x}_i,\mathbf{x}_j) 
  \ff_{j}A_{j} + \sum_{j=1}^{N_{tube}} 
  S_\varepsilon(\mathbf{x}_i,\mathbf{z}_j) \mathbf{h}_{j}\hat{A}_{j}
\]
A forward Euler method is used to evolve the positions of the helix.
We note that more rigid helices require smaller time steps than flexible
  ones in this explicit time-stepping procedure.
The positions of the two applied torques are also evolved relative to 
  the helix.
The numerical parameters used are shown in Table \ref{bigtable}.

\end{enumerate}

We remark that we also modify this algorithm  to study the swimming of a 
{\em rigid} helix 
within a tube, driven by the  torque/couter-torque actuation described above.  
In this case, given the applied torques, we need to solve for a distribution of 
  forces at the discrete points of the tube surface and those of the helix 
  surface so that (a) the fluid velocity is zero at the tube nodes and (b) 
  the velocity at the helix nodes is that of a rigid translation ${\bf {U}}$ and
  rotation ${\bf {\Omega}}$.   
The six unknowns ${\bf {U}}$, ${\bf {\Omega}}$ are determined by enforcing the 
  conditions of free-swimming (total forces and torques are zero on swimmer).
We note that this entails solving a large linear system of size 
  $3(N_{tube} + N +2) \times 3(N_{tube} + N +2)$.

\subsection{Sperm Number} \label{sec:sperm_number}
	
As in other elastohydrodynamic systems where flexible fibers are coupled to a 
  viscous, incompressible fluid, the relative importance of flow forces to
  elastic forces is an important non-dimensional parameter that governs system 
  performance (e.g. \cite{FuPowers2007, ts2007, Wandersman1}).
Following \cite{FuPowers2007}, we define the sperm number:
\begin{align}
  \label{spermnumber}
	Sp^4 = \frac{\xi^{\perp} \omega L^4}{{EI}},
\end{align}
where
\begin{align} 
	\xi^\perp &= \frac{4  \pi }{\log \left(\frac{L}{R_f}\right) + 1}.  
  \label{eq:xi_perp}\\
\end{align}
Here, $L$ is the arc length of the helical filament, $R_f$ is its 
  cross-sectional radius, $EI$ is its bending rigidity, and $\omega$ is the 
  rotational frequency achieved for the input torques. 
The perpendicular drag on the filament in free space is approximated by $\xi^\perp$
  in Eqn. \ref {eq:xi_perp}.  

Within this set-up, the macroscopic bending rigidity $EI$ of the node-spring 
  structure depends upon the individual spring constants $k_j $ and the topology
  of the spring network.
As first described in \cite{peskinlim2004} and used in \cite{Fauci4}, we can 
  precompute the $EI$ for the node-spring structure as follows: we construct a 
  straight cylindrical fiber with the same node-spring topology and same 
  individual spring constants, and then bend it into a circular arc with a 
  prescribed radius of curvature $\kappa$.  
We then compute the resulting energy in the node-spring system using 
  Eqn. \ref{eq:Energy}, arriving at: 
	\begin{align} 
	{EI} &= \frac{2{\cal E}}{\kappa^2 L}.
	\end{align}
While the flexible helical filaments we study are motivated by bacterial 
  flagellar bundles, we remark that the range of bending rigidities that we 
  examine here are at the low range of bending rigidities of single bacterial 
  flagella ($EI$ $\approx 10^{-24} - 10^{-21}$ N m$^2$) 
  \cite{pham2015,darnton2007force}.

%

\section{Results and Discussion}
\subsection{Flexible swimmer in free space} \label{freespace}

We first consider the dynamics of the  actuated model helical filament in free 
  space. 
In all of the simulations presented in this manuscript, the equilibrium helical 
  shape is fixed, as is the placement of the torque/counter torque actuation 
  (Table \ref{bigtable}).  
We will, however, vary two input parameters:  the stiffness of the springs 
  comprising the filament $k$ and the strength of the rotlet $\sigma$ that drive
  the rotation.  
We choose first to vary these separately and then analyze results in terms of 
  the non-dimensional Sperm number.

\begin{figure}
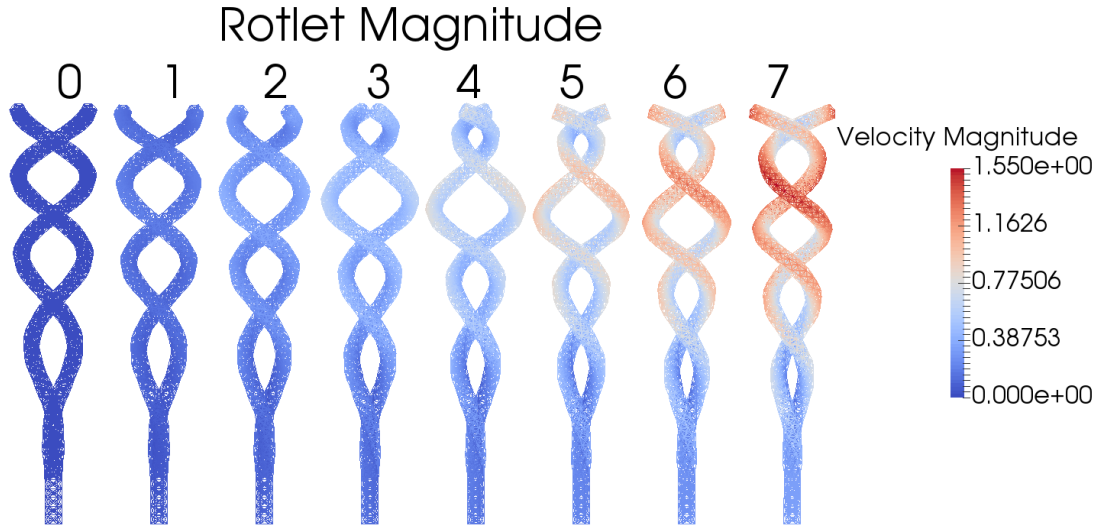

  \centering
  \includegraphics[width=0.9\textwidth]{{{flagella_labeled_by_strength}}}
  \caption{Achieved shapes of the helical flagellum for a fixed stiffness 
           constant $k=300$ and varying torque magnitude. Each image shows the 
           flagellum at two different phases for visualization purposes. The 
           leftmost image corresponds to the equilibrium configuration without 
           the dynamics. Subsequence images correspond to increasing the
           regularized rotlet magnitude by one dimensionless unit.}
  \label{fig:flagella_deformation_}
\end{figure}

Figure \ref {fig:flagella_deformation_} shows snapshots of the emergent shapes of
  the same helical filament with stiffness constant $k=300$ actuated with 
  increasing rotlet strengths. 
These snapshots are taken at times after the elastic structure has settled into 
  a steady shape.    
For each rotlet strength $\sigma = 0,1,...,7$, two projected images of the helix
  are shown.   
Note that the case of rotlet strength zero is the equilibrium configuration of 
  the helix.  
Of course, if the helix were rigid ($k = \infty$) it would maintain its 
  equilibrium shape for each rotlet strength.  
However, for this flexible filament ($k=300$), we see that stronger actuation 
  gives rise to smaller amplitude and larger wavenumbers (more turns in the 
  helix).  
These results are reminiscent of the experiments in \cite {coq2008} where a 
  flexible, natively straight filament was rotated in a viscous fluid. 
Here a shape transition to helicity was demonstrated, with smaller amplitude and
  larger wavenumbers emerging for larger rotation speeds.   

\begin{figure}
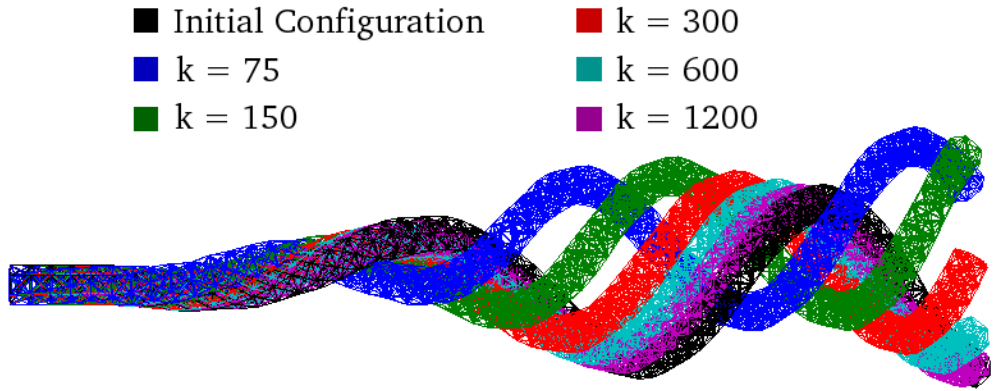

  \includegraphics[width=0.9\textwidth]{{{flagella_deformation_varying_k}}}
  \caption{Achieved shapes of the helical flagellum for a fixed rotlet magnitude 
           of 5 dimensionless units and varying spring stiffness constants. All
           images are shown at the same time of simulation.}
  \label{fig:flagella_deformation_varying_k}
\end{figure}

Figure \ref{fig:flagella_deformation_varying_k} shows snapshots of five helical 
  filaments with different stiffnesses actuated by the same rotlet strength
  $\sigma =$ 5, along with the equilibrium configuration ($k = \infty$, in 
  black). 
As in Figure \ref {fig:flagella_deformation_}, swimming progression is not shown 
  because the images are repositioned so that their front sections coincide.
This qualitative comparison of emerging shapes demonstrates that the more 
  flexible filaments actuated at the same strength exhibit larger wavenumbers.

\begin{figure}
 \label{freespacedata}
 \centering
 \subfloat[(a)]{\label{fig:speed_by_str_vary_k}\includegraphics[width=0.5\textwidth]{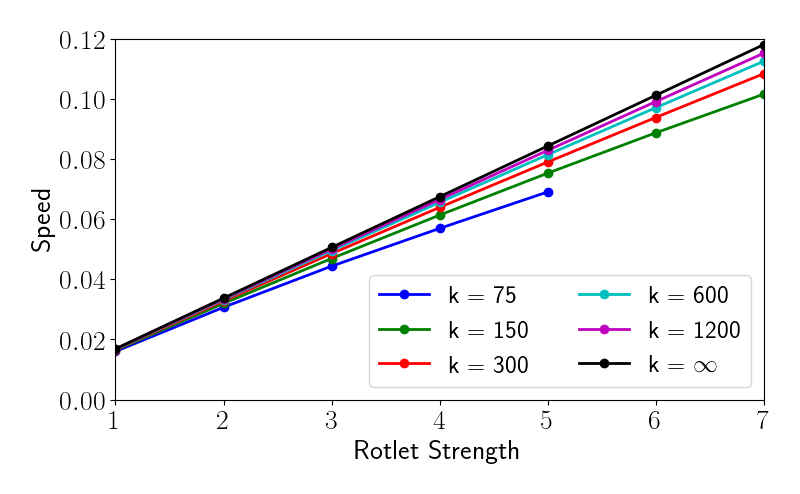}}\hfill
 \subfloat[(b)]{\label{fig:angular_vel_by_str_vary_k}\includegraphics[width=0.5\textwidth]{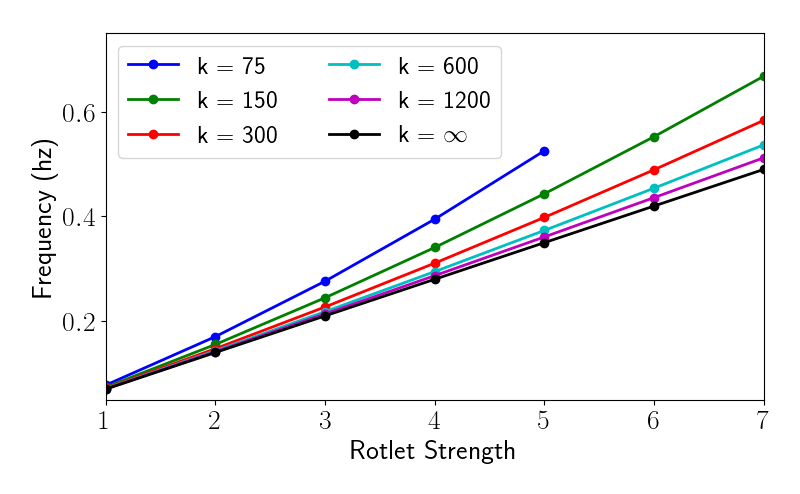}}\hfill
 \subfloat[(c)]{\label{fig:dist_by_str_vary_k}\includegraphics[width=0.5\textwidth]{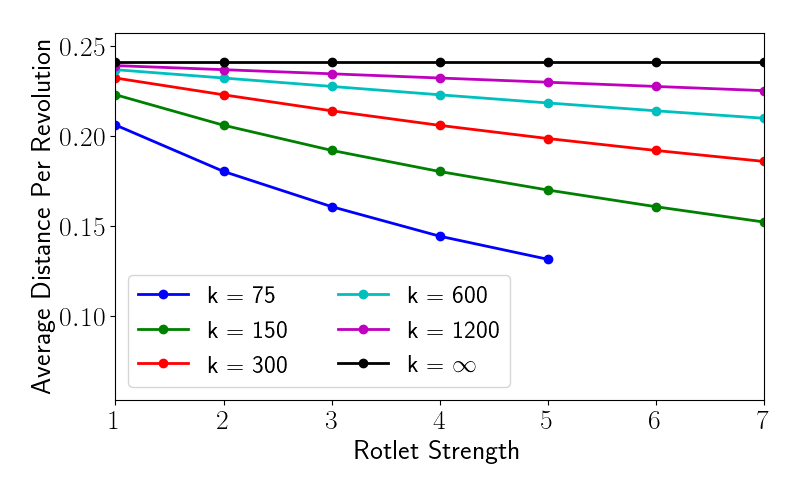}}\hfill
 \subfloat[(d)]{\label{fig:sp_dist}\includegraphics[width=0.5\textwidth]{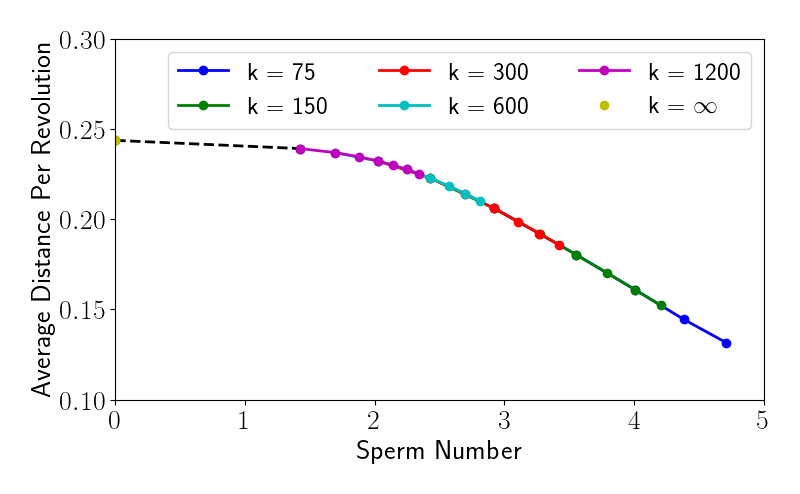}}\\
%
%
%
 \caption{(a)-(c) Swimming speed, rotational frequency, and average swimming 
          distance per revolution as a function of regularized rotlet magnitude. 
          The different curves correspond to different spring stiffness 
          constants. (d) The same data as in (c) plotted as a function of sperm 
          number. Different colors correspond to different spring stiffness 
          constants and the data points of a given color correspond to different 
          rotlet magnitudes.}
\end{figure}

Figures \ref{fig:speed_by_str_vary_k} and \ref{fig:angular_vel_by_str_vary_k} 
  show the swimming speed and the rotational frequency for each helical filament
  as a function of rotlet strength.  
For the rigid filament, both speed and rotational frequency increase linearly
   with rotlet strength, as expected.  
For a fixed rotlet strength, however, we see that swimming speed decreases with 
  flexibility while rotational frequency increases, i.e. for the same input 
  torque, flexible helices spin faster but swim more slowly. 
Figure \ref{fig:dist_by_str_vary_k} shows the distance per revolution 
  (translational distance per one spin of the helix) as a function of rotlet 
  strength. 
For the rigid helical filament, this is constant.  
As flexibility increases, the distance per revolution decreases.  

While the stiffness of the helical filament and the strength of the applied 
  torque are two things that can be controlled separately in laboratory 
  experiments (and in computational experiments),  we see that increasing 
  applied torque for a filament of a given stiffness is akin to decreasing 
  filament stiffness for a given applied torque. 
Of course, this is evident in the definition of the Sperm number $Sp$ 
  (Eqn. \ref{spermnumber}), which is a multiple of the ratio of rotational 
  frequency to filament bending rigidity.   
We remark that while we do input torque, the rotational frequency is an output 
  of the coupled fluid-filament system, so we do not know $Sp$ a priori.  
Figure \ref{fig:sp_dist} shows the distance per revolution measured for each of
  the computational simulations (six filaments of different stiffness actuated
  at seven rotlet strengths) plotted as a function of the non-dimensional Sperm
  number.  
The data in Figure \ref{fig:dist_by_str_vary_k} collapses nicely onto one curve.
For the smallest Sperm numbers, the distance per revolution is nearly constant,
  but then decreases linearly for $Sp >$ 2.5. 
                           
\begin{figure}
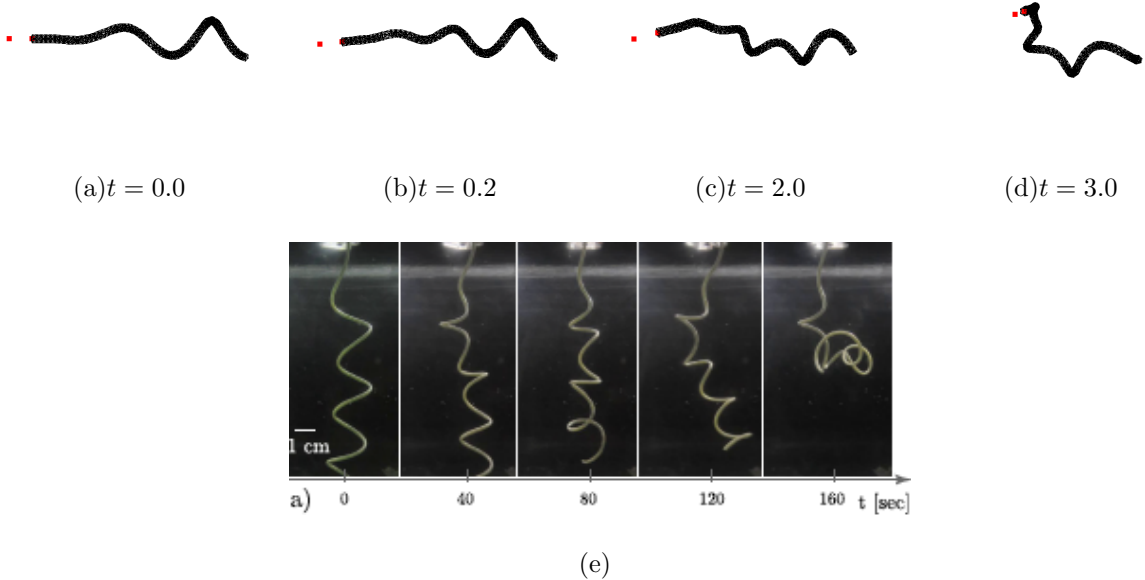

 \centering
 \subfloat[(a)$t= 0.0$]{\label{a}\includegraphics[width=0.25\textwidth]{{{buckling_t0_k75}}}}\hfill
 \subfloat[(b)$t= 0.2$]{\label{b}\includegraphics[width=0.25\textwidth]{{{buckling_t02_k75}}}}\hfill
 \subfloat[(c)$t= 2.0$]{\label{c}\includegraphics[width=0.25\textwidth]{{{buckling_t2_k75}}}}\hfill
 \subfloat[(d)$t= 3.0$]{\label{d}\includegraphics[width=0.25\textwidth]{{{buckling_t3_k75}}}}\hfill
  \subfloat[(e)]{\label{jaw}
    \includegraphics[width=0.5\textwidth]{{{spin}}}}\\
  \caption{(a-d) snapshots from a simulation with applied rotlet strength of 7 
           dimensionless units and relatively low stiffness constant $k=75$. (e)
           Buckling observed experimentally (from Jawed et al. 
           ~\cite{Jawed2015}).}
  \label{fig:buckling}
\end{figure}

Note that in Figures \ref{fig:speed_by_str_vary_k} - \ref{fig:dist_by_str_vary_k} 
  no swimming data is reported for the most flexible filament ($k=75$) actuated 
  at the largest rotlet strengths of $\sigma = 6,7$.
While all of the filaments in the other simulations relaxed into a steady shape,
  resulting in  periodic swimming motion, the most flexible filament that was 
  actuated too quickly exhibited the buckling behavior shown in 
  Figures \ref{a} - \ref{d}.
These snapshots show the time evolution of body shape, with the last frame 
  Figure \ref{d} demonstrating the total loss of a straight, helical axis.

Figure \ref{jaw} shows results of recent laboratory experiments of a rotating flexible 
  helix in a viscous fluid, exhibiting this type of buckling \cite{Jawed2015}. 
This work quantified the dynamics of the underlying mechanical instability, and 
  used both experiments and slender-body theory calculations to determine a
  critical rotational velocity $\omega _{b}$ for a given helix at which buckling 
  would occur.  
For each applied rotation $\omega$, a resulting propulsive force $\hat{F}_p$, 
  nondimensionalized as $F_p = \hat{F}_p L^2/EI$ was measured.  
Up until the critical rotational velocity $\omega _b$, $F_p$ would increase as a 
  function of $\omega $, but then the  propulsive force would drop dramatically 
  as the helix buckled.  
While the experiments using a tethered helix measure the drop-off in propulsive
  force to monitor buckling, our free-swimmer calculations show the analagous 
  drop-off in forward swimming progression when buckling occurs. 
Computational experiments have also demonstrated this buckling, called a 
  ``whirling instability", in  \cite{peskinlim2004,park2017}.    
We remark that buckling instabilities in {\em the flagellar hook} have been
  implicated as a mechanism for reorienting bacterial swimming trajectories 
  \cite{lauga2016bacterial,Nguyen2017}.

\subsection{Flexible swimmer in tube: aligned with tube axis} \label{tubealign} 

\begin{figure}
  \centering
  \subfloat[(a)]{\label{vel1}\includegraphics[width=.33\textwidth]{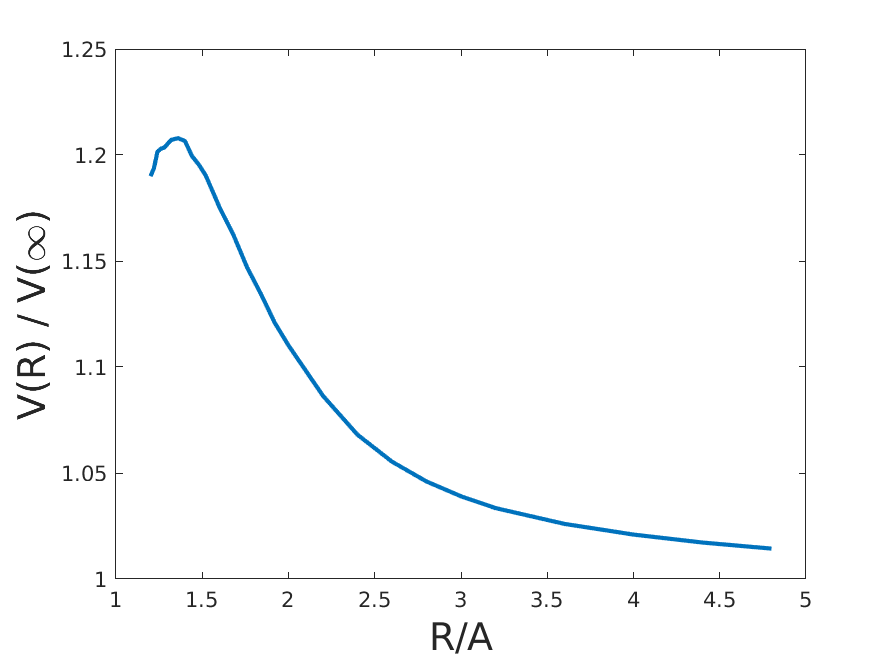}}\hfill
  \subfloat[(b)]{\label{freq1}\includegraphics[width=.33\textwidth]{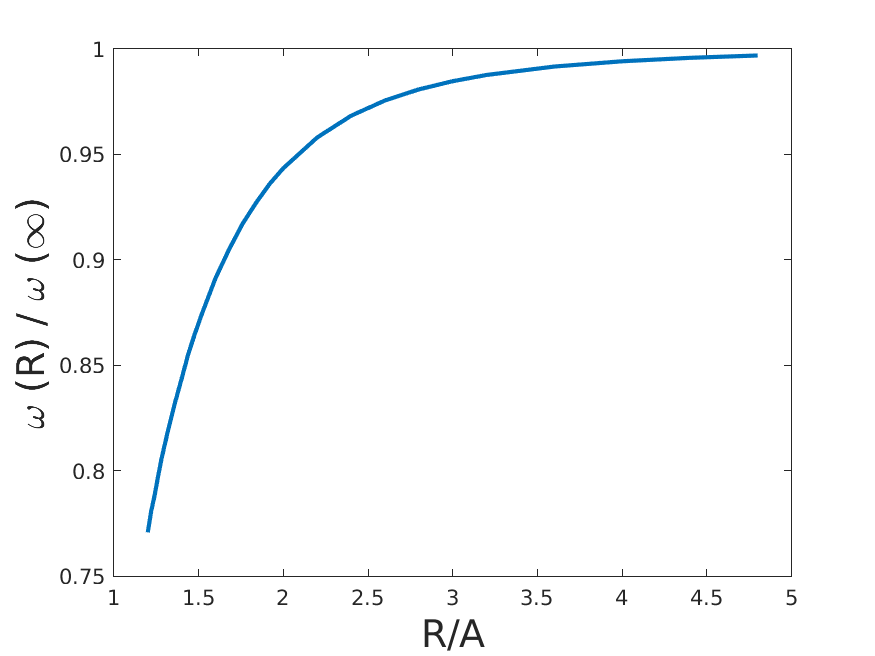}}\hfill
  \subfloat[(c)]{\label{dpr}\includegraphics[width=.33\textwidth]{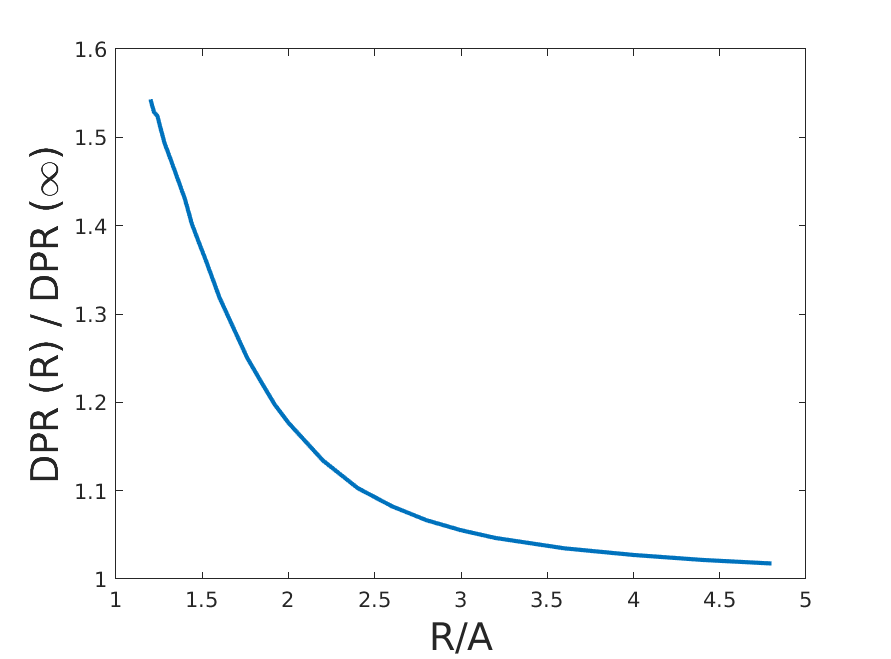}}\\
  \caption{Swimming speed, rotational frequency, and  swimming distance per 
           revolution computed as a function of tube radius for the case of a 
           {\em rigid} helical flagellum whose axis is aligned with the tube 
           axis. The tube radius is scaled by the maximum flagellum amplitude. 
           The computed quantities are scaled by their corresponding value in
           the absence of tube.}
\end{figure}

We next place our model filament inside a straight, cylindrical tube so that the
  straight axis of the helix coincides with the centerline of the tube  (as in 
  Figure \ref {fig:flagella}).
Liu et al. \cite{Liu2014} considered a related system, where a {\em rigid, 
  infinitely long} helix was driven, either by fixed torque or fixed rotational 
  velocity, to swim inside a capillary tube.  
They found that for a fixed applied torque, in all but the narrowest tubes, 
  swimming velocity increased with confinement, until the radius of the tube was
  about forty percent more than that at which the helix would touch the walls of 
  the tube.  
We first perform a series of simulations for a {\em rigid, finite} helical 
  filament with the same geometric parameters as in Table  \ref{bigtable}, 
  varying the radius of the tube $R$. 
Note that the minimum value of this radius in our simulations would be 
  $R = A + R_f$.  
Figure \ref{vel1} shows, for a fixed torque, the velocity of the helix in a 
  tube of radius $R$ normalized by its velocity in free space as a function of 
  the scaled tube radius $R/A$.  
Here we see the same non-monotonic behavior in swimming speed as a function of 
  tube radius for the finite helical swimmer as reported in \cite {Liu2014}.  
Figure \ref{freq1}, however, shows that for a fixed applied torque, the 
  rotational frequency of the helical filament decreases monotonically with 
  confinement, dropping off dramatically as the helix almost touches the tube 
  walls.  
Figure \ref{dpr} shows that the distance per revolution increases monotonically 
  with confinement - for a tightly-fitting helix the translational distance per
  turn is greatest in the tightest fits, but that turn takes a much longer time
  to complete.      

\begin{figure}
  \centering
  \subfloat[(a) ]{\label{fig:tube_speed_v_freq}\includegraphics[width=0.5\textwidth]{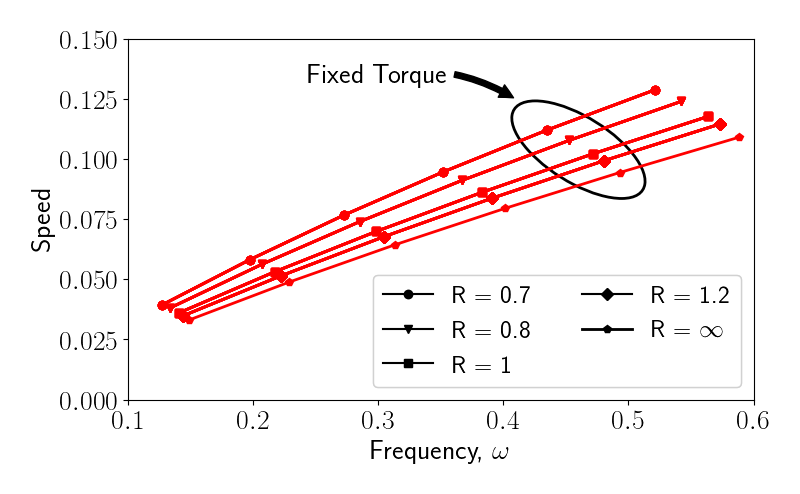}}\hfill
  \subfloat[ (b)]{\label{fig:tube_sp_dist}\includegraphics[width=0.5\textwidth]{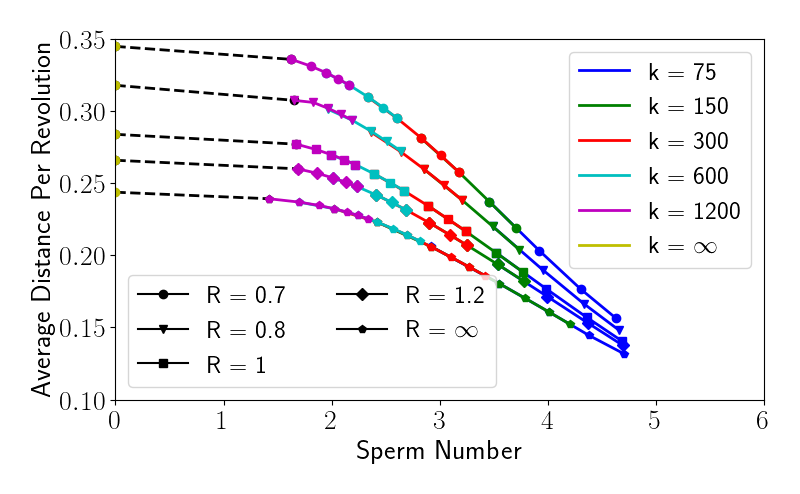}}\hfill
  \subfloat[ (c)]{\label{fig:scaled_tube_sp_dist}\includegraphics[width=0.7\textwidth]{{{tube_scaled_spd_v_dist_per_rot}}}}\\
%
  \caption{(a)  Swimming speed plotted as a function of frequency for $k = 300$ 
           (other values of $k$ are similar.) Each curve corresponds to a 
           different tube radius and the points on each curve were computed by 
           applying different torque magnitudes. 
           (b) Swimming distance per revolution plotted as a function of the 
           Sperm number. The colors indicate the varying spring stiffnesses, 
           $k$, and each curve corresponds to a different tube radius. The
           points of a given color along a curve were computed by varying the 
           rotlet strength.
           (c) Same data as in (b) but plotted as a function of the modified 
           Sperm number in a tube given in Eqn.~\eqref{Tubesperm}}
\end{figure}

We now examine the swimming dynamics of the {\em flexible} helical filaments in
  tubes of varying radii, again initialized with their axis coinciding with tube
  axis.  
Figure \ref{fig:tube_speed_v_freq} shows the swimming speed of the flexible
  helical filament with $k=300$ as a function of rotational frequency $\omega$
  in tubes of radii $ R = 0.7, 0.8, 1.0, 1.2 $ as well as in free space 
  $R = \infty$.  
Note that these simulations were performed by varying the input rotlet strength,
  and the rotational frequency is an output of the calculation.  
As expected, Figure \ref{fig:tube_speed_v_freq} indicates that for each tube 
  radius, the swimming speed increases with rotational frequency.  
We also see that the emergent rotational frequency for a fixed torque decreases 
  as the tube diameter decreases.  
Finally, for all tube radii presented in  Figure \ref {fig:tube_speed_v_freq}, 
  we see that swimming speed increases with confinement for this flexible helix.  
Here we have not included simulations in the narrowest tubes, as in Figure 
  \ref{vel1} which would show a drop-off in speed as radius decreases.

Figure \ref{fig:tube_sp_dist} shows the distance per revolution achieved by
  actuated helical filaments in a series of simulations that varied the 
  stiffness of the springs comprising the filament $k$, the applied rotlet 
  strength $\sigma$, and the tube radius $R$ as a function of $Sp$. 
Note that the curve for $R = \infty$ corresponds to Figure \ref {fig:sp_dist}.  
However, we  see that the data collapses to different curves, depending upon 
  the tube radius. 
We have already seen that the presence of the tube affects its swimming speed 
  and rotational velocity for an applied torque. 
It also affects the drag force on the helical filament. 
We estimate this effect by assuming that the tube-helix system is equivalent to 
  a helix swimming in free space with an effective drag coefficient that depends
   upon the tube radius $R$.
Let $U(R)$ be the swimming velocity of the helix in a tube of radius $R$, and 
  $C_D(R)$ be the drag coefficient. 
The drag force on the helical filament in the presence of the tube is conceived 
  as the same force resulting from motion in free space (at velocity  
  $U^\infty$) with a modified drag coefficient:
\[
  F_D = C_D(R) U(R) = ( C_D(R) f(R) ) U^\infty 
\]
Motivated by the work done in \cite{wham1996} for a sphere, we assume: 
\begin{equation}
\label{f}
  f(R) = 1 + \gamma_1 \left (\frac{R_{f}}{R} \right) + 
       \gamma_2 \left (\frac{R_{f}}{R} \right)^2 
       + \gamma_3 \left(\frac{R_{f}}{R}\right)^3
\end{equation}
so that as $R\to\infty$, we get 
  $F_D = C_D^\infty U^\infty =  \xi^\perp  U^\infty $.

Note that the drag can also be interpreted as: 
  $$F_D = C_D(R) U(R) = C_D(R) ( U^\infty f(R) )$$
  which says that the presence of the tube modifies the swimming speed by the
  same function $f(R)$.



In addition, we assume that the presence of the tube modifies the angular 
  velocity of the helical filament through a modified rotational drag 
  coefficient: $Q_D(R) w(R) = Q_D(R) ( w^\infty/g(R) )$ where: 
\begin{equation}
  \label{g}
  g(R) = 1 + \alpha_1 \left(\frac{R_f}{R}\right) 
         + \alpha_2 \left(\frac{R_f}{R}\right)^2 
         + \alpha_3 \left(\frac{R_f}{R}\right)^3.
\end{equation}
In order to estimate the coefficients in Eqs. \ref{f}--\ref{g}, we  perform a 
  least squares fit to our velocity/angular velocity data from
  simulations only in the case of the {\em rigid} helical filament:
\begin{align}
\gamma_1 = 0.3371, \qquad \gamma_2 = -1.7293, \qquad \gamma_3 = 1.2798,
\end{align}
and
\begin{align}
\alpha_1 = -0.0172, \qquad \alpha_2 = 0.0092, \qquad \alpha_3 = 0.5261.
\end{align}
The distance per rotation of the helical filament is:
\begin{equation}
\frac{U(R)}{w(R)} = \frac {(f(R)g(R)) U^\infty}{w^\infty},
\end{equation}
and we thus define a scaled distance per revolution by dividing by the values 
  plotted in Figure \ref {fig:tube_sp_dist} by  $f(R)g(R)$.

Similarly, the modified sperm number is:         
\begin{equation}
\label{Tubesperm}
  Sp(R)^4 = \frac{ C_D(R)w(R) L^4 }{EI} 
          = \frac{ C_D^\infty w^\infty L^4 }{f(R)g(R) EI}
          =  \frac{Sp^4_\infty}{f(R)g(R)}.
\end{equation}
        %
%
Figure \ref{fig:scaled_tube_sp_dist}, shows the {\em scaled} distance per 
  revolution as a function of the modified sperm number $Sp(R)$ for all of the 
  data points shown in Figure \ref{fig:tube_sp_dist}.
We see that all the data collapses onto the free space curve indicating that the 
  two scalings we have introduced appropriately capture much of the dynamics due
  to the tube.
We also emphasize that the evaluation of the coefficients in the expansion of 
  the functions $f(R), g(R)$ used only the simulation data for the {\em  rigid}
  helical filaments in tubes of varying radii. 
The collapse of all data, including data from the simulations of the flexible
   helical filaments,  is a good indication that the influence of the tube
   appears mostly in the form of drag--like forces and not as a consequence of 
   deformations of the filaments.
However, we see that there is greater variation in the scaled distance per 
  revolution associated with larger $Sp$, because the changes in shape as a 
  result of the confinement are more pronounced for the more flexible swimmers.

\subsection{Flexible swimmer in tube: not aligned with tube axis} 

\begin{figure}[h!tp]
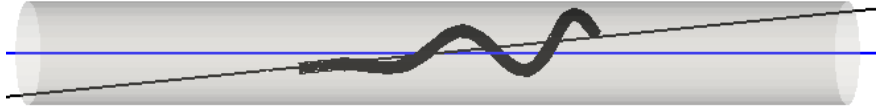

  \centering
  \includegraphics[width=0.7\textwidth]{{{angled_flagella_in_tube}}}
  \caption{Initial orientation of a flagellum angled $0.1$ radians away from the 
           central axis of a tube. The centerline of the tube is indicated by 
           the blue line and the black line is the centerline of the targeted 
           configuration of the filament.}
  \label{fig:angled_flagella_in_tube}
\end{figure}

In the simulations presented above, where the swimmer was initially launched so 
  that its axis  coincided with the axis of the tube, the actuated swimmer
   continued to remain centered.  
Here we examine how the swimming trajectory would change if the initial position 
  of the helix were not aligned with the tube axis as in 
  Figure \ref {fig:angled_flagella_in_tube}.  
Would the swimmer eventually hit the wall?  
Would it straighten out its path to swim down the center of the tube?  
Moreover, how does this depend upon the radius of the tube?   
Here we present three simulations for a flexible swimmer with $k=300$ driven by
   a rotlet strength of $\sigma= 5$ inside of tubes of radii 
   $R = $ 0.675, 0.725, 0.775.  

\begin{figure}
\centering
  \subfloat[Time ]{\label{times}\includegraphics[height=8.75cm, width= 0.08 \textwidth]{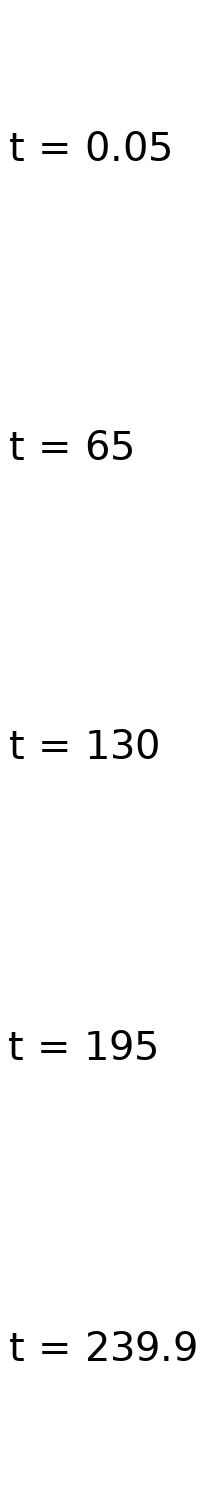}}\hfill
  \subfloat[(a)$R=0.675$ ]{\label{one}  \includegraphics[height=8.75cm, width= 0.2928\textwidth]{{{flagella_panel_r=0675}}}}\hfill
  \subfloat[(b)$R=0.725$ ]{\label{two}  \includegraphics[height=8.75cm, width= 0.2928\textwidth]{{{flagella_panel_r=0725}}}}\hfill
  \subfloat[(c)$R=0.775$ ]{\label{three}  \includegraphics[height=8.75cm, width= 0.2928\textwidth]{{{flagella_panel_r=0775}}}}\\
  \caption{Side view of the flagellum in three tubes  of different radii. Each 
           column shows five snapshots corresponding to a particular tube 
           radius. Each filament was initially angled 0.1 radians away from the 
           central axis of a tube.}
  \label{swimmercenter}
\end{figure}

Figure \ref{swimmercenter} shows a sequence of snapshots in time of the swimmer 
  in tubes of increasing radii, where the initial orientation of the helical 
  filament formed a non-zero angle with the tube centerlines as in 
  Figure \ref {fig:angled_flagella_in_tube}.  
The forward progression is suppressed in these projected images.  
In each of the three simulations, we see that the helical shape achieved is 
  basically the same for all three radii.  
However, we see that in the largest tube $R = $ 0.75 (panel (c)) the angle 
  between the horizontal tube axis and the helix axis has the greatest 
  variation, while this angle in the smallest tube $R = $0.675 appears to be 
  approaching zero (panel (a)).

\begin{figure}
  \centering
  \subfloat[Time ]{\label{times}\includegraphics[height=8.75cm, width= 0.08 \textwidth]{{times}}}\hfill
  \subfloat[(a)$R=0.675$ ]{\label{one1}  \includegraphics[height=8.75cm, width= 0.228\textwidth]{{{combined_2d_panel_r=0675}}}}\hfill
  \subfloat[(b)$R=0.725$ ]{\label{two2}  \includegraphics[height=8.75cm, width= 0.228\textwidth]{{{combined_2d_panel_r=0725}}}}\hfill
  \subfloat[(c)$R=0.775$ ]{\label{three3}  \includegraphics[height=8.75cm, width= 0.228\textwidth]{{{combined_2d_panel_r=0775}}}}\hfill
  \subfloat[(d)$R=0.675$ ]{\label{helical1}\includegraphics[width= 0.33 \textwidth]{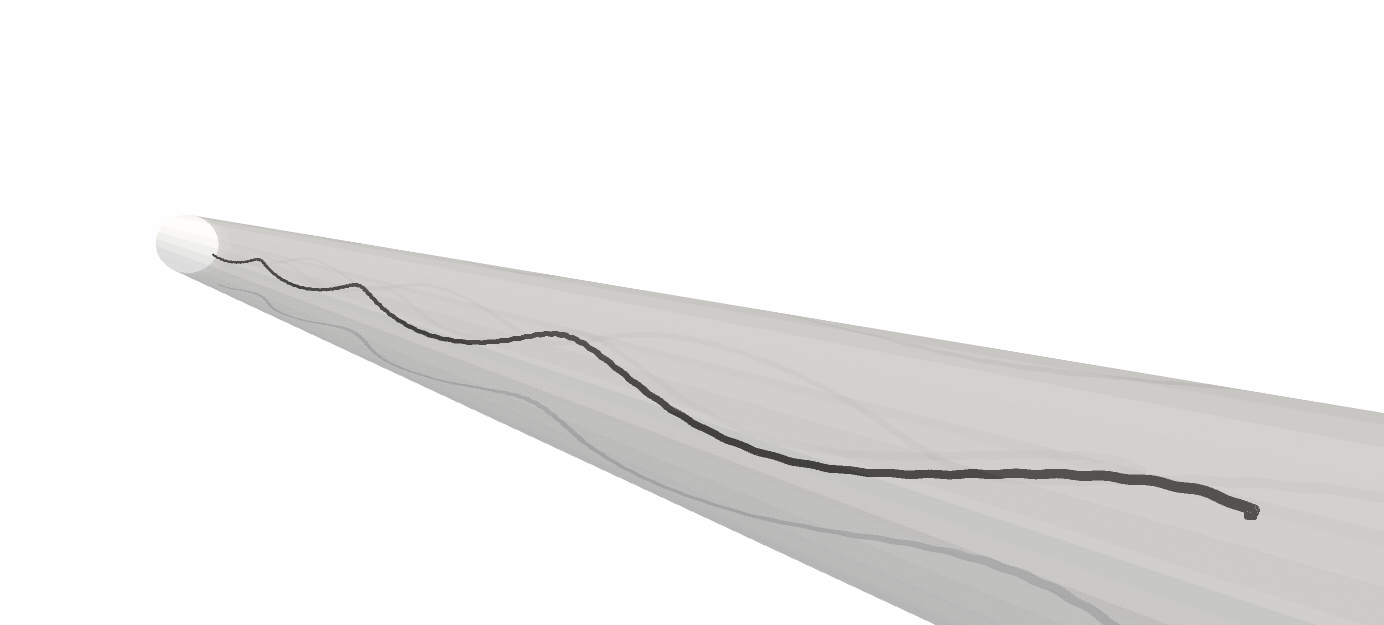}}\hfill
  \subfloat[(e)$R=0.725$ ]{\label{helical2}\includegraphics[width= 0.33 \textwidth]{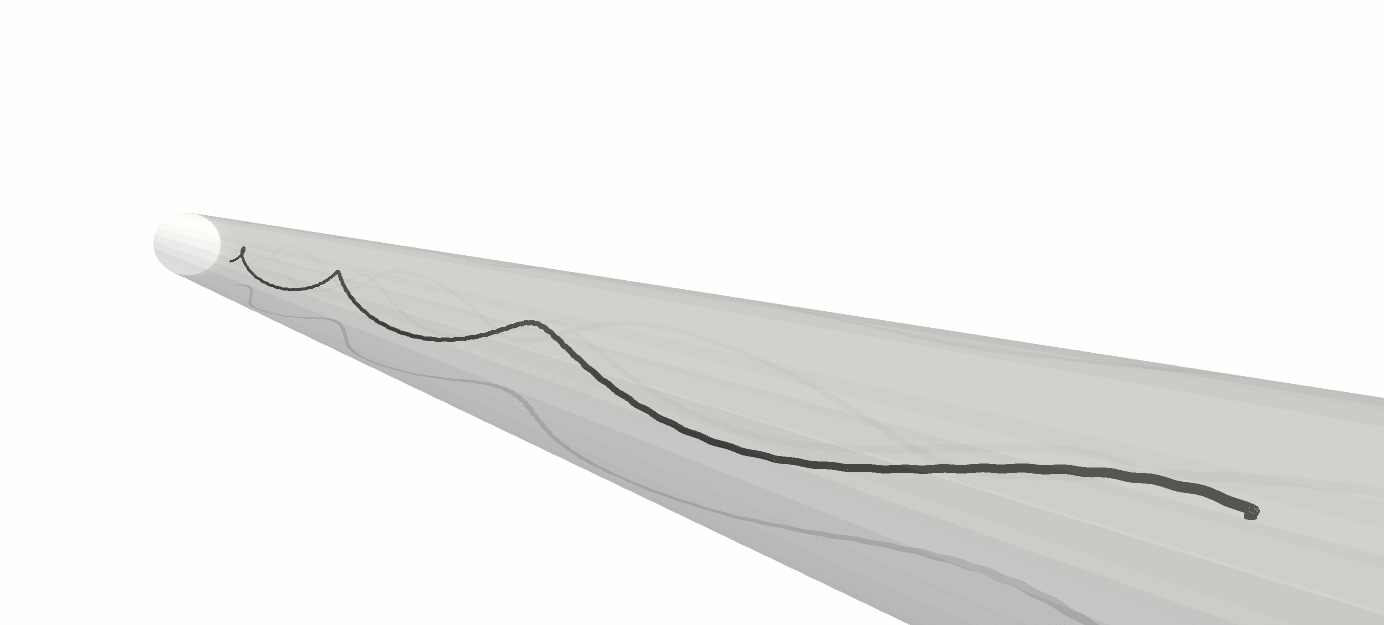}}\hfill
  \subfloat[(f)$R=0.775$ ]{\label{helical3}\includegraphics[width= 0.33 \textwidth]{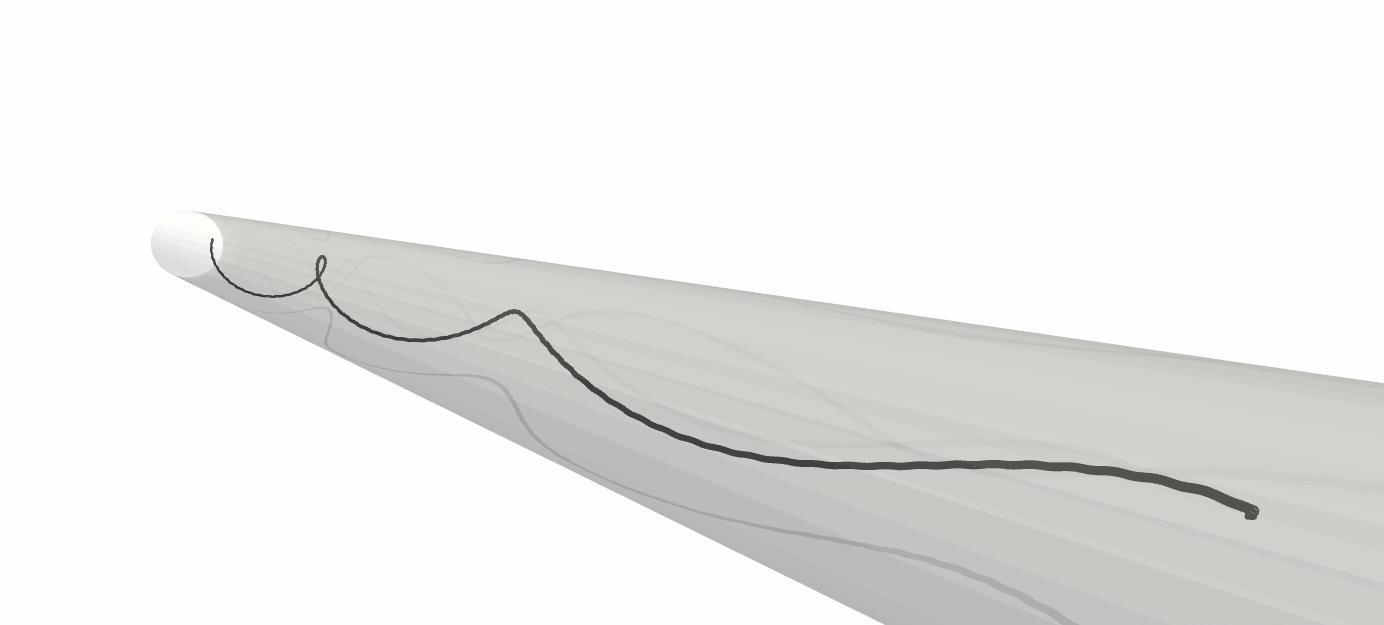}}\\
  \caption{Axial view of the snapshots in Figure~\ref{swimmercenter}. Each 
           flagellum was initially angled 0.1 radians away from the central axis
           of a tube.
           (a) Flagellum and trajectory of the centerpoint of the front 
               cross-section for a tube radius of $R=0.675$.
           (b) Flagellum and trajectory of the centerpoint of the front
               cross-section for a tube radius of $R=0.725$.
           (c) Flagellum and trajectory of the centerpoint of the front 
               cross-section for a tube radius of $R=0.775$.
           (d)-(f) Perspective views of the trajectory of the front point as the
                   flagellum swims down the tube.}
  \label{figlast}
\end{figure}

We present another view of the swimming progression in these three simulations 
  in Figure \ref{figlast}.
For each tube radius, there are two columns from the perspective of looking 
  directly down the tube axis. 
In the right-hand-side column, we see the projection of the helical swimmer com
  ing toward the reader.  
In the left-hand-side column, we see the projected trajectory of the position 
  of the countertorque, at time zero indicated by a star.  
Note that if the swimmer were aligned with the tube axis, this projected
   trajectory would be a single point.  
Figures \ref {figlast}(d-f) give three-dimensional views of the trajectories of
  the swimmers, this time looking down the tube as the swimmer moves away from
  the reader.  
We see that in the smallest tube ($R = $ 0.675), the swimmer exhibits centering 
  behavior, with a helical trajectory whose radius is getting smaller with time.  
In the largest tube ($R = $ 0.775), the swimmer's helical trajectory carries it 
  towards the tube walls, and, in fact, because we are not including repulsion
  in this model, the simulation is suspended when the swimmer hits the wall. 
Intriguing behavior is seen in the middle sized tube ($R = $ 0.725), where the
  swimmer settles upon a limit cycle such that it rolls around the tube in a 
  periodic manner, as can be seen by the projected circular trajectory of the 
  counter-rotlet in Figure \ref {figlast}(b), and the three-dimensional helical
  trajectory in Figure \ref {figlast}(e).

The above examples show that when the initial orientation of the swimmer is
   perturbed slightly from alignment with the axis of the tube, three classes of
   swimming trajectories emerge.  
These classes of swimming trajectories are consistent with the swimming 
  trajectories computed by Shum and Gaffney \cite {ShumChannel} for bacterial 
  cells swimming between two planar boundaries.   
For large plate gaps, cells with long enough helical flagella were attracted to 
  the wall, while for very narrow gaps, the swimmer relaxed to a trajectory
   midway between the walls.  
At some intermediate spacing, the swimmer ``bounced" repeatedly from wall to wall.


\section{Conclusions}
In summary, we have considered the swimming of a flexible helix both in free 
  space and in a capillary tube.  
When driven by a fixed torque/counter-torque system, the swimming velocity of 
  the helical filament decreases with flexibility, but its rotational velocity 
  increases.  
We find that the swimming performance, when measured by distance traveled per 
  revolution, is well-described by the Sperm number. 
We have also demonstrated that buckling of the filament occurs for the most
  flexible helices actuated with large rotlet strengths ($Sp > 4.7 $).

For the same flexible helices driven by a fixed torque/counter-torque swimming 
  along the centerline of a tube, we find that the swimming speed increases with
  confinement, as does the distance traveled per revolution.  
This enhanced swimming performance decreases with helix flexibility.  
Using a modified Sperm number that accounts for the surface of the tube's effect 
  on drag coefficients, we again find that swimming performance in the tube can
  be well-described by this non-dimensional parameter.  

When the alignment of the swimmer is perturbed from the tube axis, we find that
  for larger tubes, the swimmer will eventually hit the boundary of the tube.  
However, for tubes of smaller diameter, the helical swimmer actually centers 
  itself to align with the tube axis.  
This finding, along with similar results for swimmers between planar boundaries
  \cite{ShumChannel} suggest the provocative idea that bacterial cells may have 
  an easier time breaking through a tightly-packed porous medium with small 
  pores.  
Bacteria moving through large pores are likely to adhere to the matrix.

\section{Acknowledgements}
This research was supported by 
the National Science Foundation grant DMS-1043626 and the Gulf of Mexico 
  Research Initiative.
The authors would like to thank Kyriakos Papadopoulos and Mengyuan Zheng for 
  helpful discussions.

\clearpage
\bibliographystyle{plain}
\bibliography{citations}

\end{document}